\documentstyle[preprint,revtex]{aps}
\begin{document}
\draft
\begin{title}
\bf
Dynamics of Inhomogeneous Tomonaga-Luttinger \\ Liquid Wire
\end{title}
\author{V.V.Ponomarenko}
\begin{instit}
Frontier Research Program, RIKEN, Wako-Shi, Saitama
351-01, JAPAN.  \\
Permanent address: A.F.Ioffe Physical Technical Institute,\\
194021, St. Petersburg, Russia.
\end{instit}

\date{\today}

\begin{abstract}
Dynamics of the 1D electron transport between two reservoirs
 are studied  based on  the inhomogeneous Tomonaga- Luttinger Liquid (ITLL)
model
 in the case when the effect of the electron backscattering on the impurities
is
negligible. The inhomogeneities
of the interaction lead to a charge wave reflection.
This effect supposes a special behavior of the transport
characteristics at the microwave frequencies.
New features are predicted in the current noise spectrum and in the a.c.
current-
frequency dependence.
 Knowledge of them
may be very useful to identify experimental setup with the one specified
by the ITLL model.

\end{abstract}


\widetext
The effect of the electron-electron interaction has been suggested to be
especially
important for 1D electron transport through a few channel wire \cite{KF}.
Such a wire is generally believed in view of the results of the 1D field theory
\cite{Fr}
to be described as a Tomonaga-Luttinger Liquid.  A number of authors
\cite{Number} developed this approach making it closer to experimental reality.
Recently a new modification has
been suggested \cite{M},\cite{SS}, \cite{me} to cope with the effect of a
finite length of the quantum wire.
It supposes that the 1D transport through the
quantum wire of finite length may be described by the Tomonaga-
Luttinger Liquid model with an inhomogeneous constant of the interaction.
The model meets a natural demand of the unrenormalizableness
of the zero-frequency conductance by the interaction. It has also
found agreement with a recent experiment \cite{T}
where the tracks of the interaction effect predicted before \cite{OF}
were allegedly observed. Therefore it looks tempting to search for
further consequences of this model which could be useful in comparing with
the experimental results.

This paper is aimed at analyzing the simplest dynamical features of the 1D
transport following from the inhomogeneous Tomonaga- Luttinger Liquid (ITLL)
model
in the case
when the effect of the electron backscattering on the impurities is
negligible. Such a situation seems to be easily reachable via proper
tuning of the gate voltage \cite{T}, governing the depth of the potential
well forming the wire. In this system an electron travels from one
reservoir to the other without reflection. The inhomogeneities
of the interaction, however, vary velocity and charge of the
charge wave excitations, what may be interpreted as a charge wave reflection
\cite{SS}.
This effect can lead to a special behavior of the transport
characteristics at microwave frequencies. To study this behavior I will
impose two restriction on the transport experiment in the two probe
geometry
in addition to the above condition of the absence of
the backscattering and  the adiabatical joint with the reservoirs:
i) the interaction between electrons in the wire is local since a gate
electrode located near the wire screens the Coulomb interaction between
the wire electrons at the distances more than
the range of the order of the width of the wire
which is much less than the length $L$ of the wire;
ii) current entering a reservoir from the 1D conductor is immediatly
transfered to the attached lead whereupon the current operator evaluated
at the end of the 1D wire will give the experimentally observed value.
This is valid
as charge redistribution in the leads is
of a much higher frequency than the frequencies that will be discussed.
Having in mind the GaAs structure one can reckon that the width of the wire
is several hundred angstroms, its length is about 2 $\mu m$, the Fermi velocity
$v$ is about $10^{-5}$ m/sec, and the frequency, corresponding to the
traversal time $t_L$, is $ \approx 1K$.

The transport through such a system with the one spin-degenerate channel
wire can be described,
following \cite{me}, and also \cite{M}, \cite{SS},
with the Lagrangian ($\hbar = 1$)
\begin{eqnarray}
{\cal L}_t={v \over{8 \pi}} \int dx [ \frac{1}{v^2}(\partial_t \phi(t,x))^2
- u^2(x)(\partial_x \phi(x,t))^2 ] +
\frac{e}{\sqrt{2} \pi} \int dy \phi(t,y) \partial_y V(t,y)  \label{1} \\
u^2(x)=(1+2 \varphi(x) U/(v \pi)) \nonumber
\end{eqnarray}
The bosonic field $\phi$ corresponding to the charge wave excitations are
connected with the operator of the electron density as $ \rho =
\frac{1}{\sqrt{2} \pi} \partial_x \phi $ and with the operator of the electron
current as $ j = - \frac{e}{\sqrt{2} \pi} \partial_t \phi $
due to a bosonization procedure \cite{Fr} . $v$ is the Fermi
velocity, and the local density-density interaction $\varphi(x) U$
between the electrons is switched on in between the points $x=0$ and $x=L$,
where the channel adiabatically joins the reservoirs.

The second part of the Lagrangian includes the applied electric field
equal to $ - \partial_y V(t,y) $. In general, it makes the problem a non-
equilibrium one. Following a standard procedure \cite{NE} one should
calculate any physical quantity as a functional average of the appropriate
function of $\phi$ with the $exp\{ i{\cal S}( \phi )\}$ weight,
where the action
$S( \phi )$ is the integral of ${\cal L}_t$ over Schwinger's contour
from $-\infty$ to $\infty$ and back. With the Gaussian form of the Lagrangian
the second term in (\ref{1}) may be excluded due to shifting of $\phi$ by
$<\phi (t,x)>= (1/v)\int dt' \int dy G(t-t',x,y) (-\partial_y V(t',y))$. Here
$G(t,x,y)$ is the retarded Green function, whose Fourier transform
satisfies
\begin{equation}
\left[  \frac{\omega^2}{v^2} +\partial_x u^2(x) \partial_x \right]
G(x,y, \omega )=\delta(x-y)
\label{2}
\end{equation}
After the shift of $\phi$ all calculations can be done with the equilibrium
Lagrangian.

In this way of calculation one can find that the current $<j(t,x)>$ is always
linearly connected with the voltage in this model
(it is not surprising in view of the linearized dispersion law of
the TLL model),
and the conductivity per
spin channel is $\sigma(x,y,\omega )=\frac{e^2}{\pi v} i\omega G(x,y, \omega
)$.
Another useful consequence is that the current-current correlator at the steady
voltage distribution
\begin{equation}
P(x,\omega)=\int dte^{i \omega t} ( \frac{1}{2}<[j(x,t),j(x,0)]_+>-<j(x,0)>^2)=
2 coth(\frac{\beta \omega }{2})
 \omega Re \sigma(x,x,\omega)
\label{3}
\end{equation}
does not depend on the voltage. Hence, it can be represented due to
the fluctuation-dissipation theorem as the real part of the conductivity
$\sigma(x,x,\omega)$. $\beta=T^{-1}$ is the inverse temperature.

The conductivity may
be constructed from the two independent solutions
of the homogenious Eq. (\ref{2}), corresponding to a charge wave coming from
the left(right) side
$ f_{\pm \omega}(x)= t(\omega ) exp{ \pm i(\omega/v) x}$ at $x \rightarrow \pm
\infty , ( t(\omega )=t(- \omega) $ are the amplitudes of the transmission )
as \cite{me}
\begin{equation}
\sigma (x,y, \omega )=\frac{e^2}{2 \pi t(\omega)}
[f_{+ \omega}(x)f_{- \omega}(y) \theta(x-y)
+f_{+ \omega}(y)f_{- \omega}(x) \theta(y-x) ] \\
\label{4}
\end{equation}
In the limit $ \omega \rightarrow 0 $ the conductivity becomes constant
\cite{M}, \cite{SS}, \cite{me}
equal to the universal conductance $ \sigma_0=e^2/(2 \pi ) $
since the only restricted
solution of the homogeneous Eq. (\ref{2}) at $\omega =0$ is constant.

For further consideration it is important that $u(x)$ may be approached
by a step function as
only excitation of the waves of the electron density with small $k$ (in
comparison with the inverse width of the channel and with
the inverse distance to the screening gate, but not with the inverse length
of the wire $L^{-1}$ ) is essential. In this case the solutions of the
homogeneous Eq. (\ref{2}) take the form
\begin{eqnarray}
f_{+ \omega}(x) = \left\{ \matrix{e^{ix \omega/v }+r(\omega)
e^{-i x\omega/v },  & x<0 \cr
t(\omega )e^{i L\omega/v }
[a_+ e^{i (x-L)\omega/v'} + a_- e^{-i (x-L)\omega/v'}],  &0<x<L \cr
t(\omega )e^{i x\omega/v },  &L<x \cr} \right. \label{5} \\
f_{- \omega}(x) = \left\{ \matrix{
t(\omega) e^{-i x\omega/v },  & x<0 \cr
t(\omega )[a_+ e^{-ix \omega/v'} + a_- e^{i x\omega/v' }],  &0<x<L \cr
 e^{-i x\omega/v }+
r(\omega )e^{i(x-2L) \omega/v },  &L<x \cr} \right.
\nonumber \\
t(\omega )=\frac{e^{-i L\omega /v}}{cos(\omega t_L)-i \frac{1}{2}
(\frac{1}{u}+u)sin(\omega t_L)}, \ \ \ r(\omega)=
\frac{-i \frac{1}{2}(\frac{1}{u}-u)sin(\omega t_L)}
{cos(\omega t_L)-i \frac{1}{2}(\frac{1}{u}+u)sin(\omega t_L)}
\label{6}
\end{eqnarray}
where $a_{\pm }=(1 \pm u^{-1})/2, u$ is the meaning of the function $u(x)$
inside the wire $(u=u(L/2))$, $v'=vu$ and  $t_L=L/v'$
are the velocity and the traveling time of the elementary charge excitation
of the wire.
First let me calculate the current noise spectra for the current coming
into the right reservoir $J_R(t)=j(L+0,t)$ and for the current outgoing
from the left reservoir $J_L(t)=j(-0,t)$. In fact, both spectra coincide
and are equal to
\begin{equation}
P(-0,\omega)=P(L+0,\omega)=\sigma_0 \omega \left(1- \frac{1}{4}
\frac{(u^2-u^{-2})sin^2(\omega t_L)}{cos^2(\omega t_L)+(u+u^{-1})^2/4
sin^2(\omega t_L)}\right)
\label{7}
\end{equation}
for $1/\beta=T \ll \omega \approx 1/t_L$. This result shows  (fig.1)
that a linear dependence of the free electron noise
spectrum \cite{Ya} is modulated by the
oscillating factor with the period equal to $ \pi/t_L $.

Another way to probe dynamics of the transport at the frequencies
$ \geq \pi/t_L $
is making use of a microwave electro-magnetic radiation. Below I will
examine its effect on the frequency dependence of the amplitude
of the left and right lead currents. The radiation acts on the system
by varying both the chemical potentials and by accelerating the charge waves
inside the 1D wire.  The resultant effect can be specified with the
effective electro-chemical potential as
\begin{equation}
-\partial_x V(x,t)=cos(\omega t) [V_L \delta(x)+ V_R \delta(x-L)
+E \theta(x) \theta (L-x)]
\label{8}
\end{equation}
where $\delta(x)$ and $\theta(x)$ denote Dirac's delta and
Heaviside's step functions, respectively. Since these three
terms have a different origin I first will consider them separately.

The $V_L$ term in (\ref{8}) pertains to such an experiment when
the chemical potential of the right lead is fixed, say $\mu_R=0$, and the
chemical potential of the left lead is driven by the microwave
radiation imposed between the left lead and the ground electrode
somewhere far from the wire. Then the amplitudes of the right $J_{Rc}$ and
left $J_{Lc}$ currents may be calculated as
\begin{eqnarray}
J_{Rc}=2V_L \sigma_0 |t(\omega)|=\frac{ 2V_L \sigma_0}{\sqrt{
cos^2(\omega t_L)+(u+1/u)^2/4 sin^2(\omega t_L)}} \nonumber \\
J_{Lc}=2V_L \sigma_0 |1+r(\omega)|=2V_L \sigma_0 \sqrt{2-
\frac{1+(u^2-1/u^2)/2 sin^2(\omega t_L)}
{cos^2(\omega t_L)+(u+1/u)^2/4 sin^2(\omega t_L)}}
\label{9}
\end{eqnarray}
It reveals that both amplitudes oscillate  around the constant signal of
their free electron behavior (Fig.1) with the period $\pi/t_L$.
Their variations
are characterized by the declination of their minimums, reached at
$\omega$ equal to $ half integer \times \pi/t_L $, from $2V_L \sigma_0 $.
The minimums are $4V_L \sigma_0/(u +1/u)$ and $4V_L \sigma_0/(u +1/u)/u$
for the right and left lead current, respectively. Fig.1 depicts
dependence of the normalized amplitudes $J_{L(R)c}/(2V_L\sigma_0)$ on the
frequency. It shows that for the small interaction ($u=1.43$ was taken
to accord with \cite{T}) the normalized amplitude of
the left lead current ( full line) turns out to be very close to
the normalized noise signal
$P(\omega )/(2\sigma_0 coth(\beta \omega/ 2))$ (dotted line).

The above oscillations results from the scattering of the charge waves
on the places of the interaction inhomogeneity. It is better
seen from the time dependence of $J_R(t)$ produced by a sudden jump of the
chemical potential of the left reservoir, what corresponds to the switching
on of the voltage as $ V_L(t)=V_L \theta(t)$. This current reply may be
calculated \cite{SS} as
\begin{equation}
J_R(t)=2 V_L \sigma_0 i \int \frac{d \omega}{2 \pi} \frac{t(\omega)
e^{i \omega L/v}}{\omega+i0}=
2 V_L \sigma_0 \theta(t)\left\{1-\left(\frac{u-1}{u+1}\right)^
{2int[(t/t_L+1)/2]}
\right\}
\label{10}
\end{equation}
where $int[x]$ equals the integer part of $x$, and $(u-1)/(u+1)$ is the
reflection coefficient. The first step of the current-time staircase
dependence of Eq. (\ref{10})
appears for $t \ge t_L$, the second one  for $t \ge 3t_L$ and so on.

Another experimental situation arises when the radiation acts on the
electrons only inside the wire. Surmising that the electric field $E$
parallel to the channel
is uniformly distributed along the length $L$, one can write the right
lead current as $J_R=2E|\int_0^L dy \sigma(L,y,\omega)|
cos(\omega t-arg \{\int_0^L dy \sigma(L,y,\omega)\} )$. Its amplitude can be
found as
\begin{equation}
\frac{J_{Rc}}{2EL \sigma_0}=\frac{sin(\omega t_L/2)}{\omega t_L/2}
\sqrt{\frac{1}{2} \left( 1+\frac{1}{u^2} \right)}
\sqrt{\frac{1+\frac{u^2-1}{u^2+1}cos(\omega t_L)}{cos^2(\omega t_L)
+(u+1/u)^2/4 sin^2(\omega t_L)}}
\label{11}
\end{equation}
The first multiplier on the right side (\ref{11}) displays the signal
in the free
electron case ( $t_L$ is renormalized by the interaction ). It becomes
$\pi/t_L$ periodically modulated by the interaction term.
The whole signal quickly vanishes with increase of the
frequency (Fig.2).

Turning to a general situation one can represent the amplitude of the
current coming into the right electrode
as
\begin{equation}
J_{Rc}(\omega )=2 \sigma_0 |V_L t(\omega )e^{iL\omega /v} +V_R(1+r(\omega ))
+E\frac{iv}{\omega }(1-r(\omega )-t(\omega )e^{iL\omega /v})|
\label{12}
\end{equation}
This expression displays a ravelled dependence of the amplitude
on the frequency (Fig.2).
Simplification comes, however, with increase of
the frequency, where only abrupt steps of the electro-chemical potential
near the edges of the electrodes contribute substantially.
Finally, Eq. (\ref{12})
at $\omega t_L \gg 1$ takes the form
\begin{equation}
\left(\frac{J_{Rc}(\omega )}{2\sigma_0} \right)^2=
2V_R^2+\frac{2V_L V_R cos(\omega t_L)+V_L^2-V_R^2-1/2 V_R^2(u^2-1/u^2)
sin^2(\omega t_L)}
{cos^2(\omega t_L)+(u+1/u)^2/4 sin^2(\omega t_L)}
\label{13}
\end{equation}
The $2\pi/t_L$ periodical oscillations already exist in the free electron
case if both chemical potentials of the left and right electrodes are
rotated simultaneously. In the interacting wire the $\pi/t_L$ periodical
oscillations additionally appear, which deform the simple $cos$ law behavior
of the free electron dependence. The case displayed in Fig.2 corresponds
to the asymmetrical distribution of the electro-chemical potential
( $V_L/V=0.15, V_R/V=0.35$, where $V$ is the whole variation).

In summary, the effect of the inhomogeneities of the electron-electron
interaction on the current noise spctrum and on the a.c. current amplitude
was identified in the microwave frequency range. The current noise
spectrum turns out to be modulated by the oscillating factor with
period $\pi/t_L $ due to the inhomogeneous interaction. The a.c.current
behavior depends on the distribution of the electro-chemical potential
produced by the microwave radiation. In the case of a non-zero $V_L$ and/or
$V_R$ the amplitude oscillates in a wide range of the frequency with the
period equal to $2 \pi/t_L$ in the general case and  to $ \pi/t_L$ if
$V_L V_R=0$. These signals give a way to identify the effect of the
interaction in the experiment on 1D electron transport. The microwave
field accelerating
electrons along the 1D conductor results in the signal quickly disappearing
with increase of the frequency. Extraction of the $ \pi/t_L$ oscillating
fraction of it looks especially challenging.

I am very grateful to K. K. Likharev for
elucidative discussions and to M. Stopa for useful comments and kind help.
The interest of A. Kawabata and S. Tarucha
 is much appreciated. This work was supported by the STA.

\figure{Dependences of the normalized spectrum of the current noise
(curve 1, dotted) and of the
amplitudes of the right (curve 2, dash-double-dot) and left
(curve 3, full) lead current for the inetraction parameter
$u=1.43$ corresponding to the experiment \cite{T}.
The spectrum $P(\omega )$ is normalized as
$P(\omega )/(2\sigma_0 coth(\beta \omega/ 2)), \sigma_0=e^2/(2\pi )$, where
$\beta$ is the inverse temperature. The normalization of both current
amplitudes $J_{R(L)c}$ was suggested $J_c/(2V_L \sigma_0 )$,
where $V_L$ is the
amplitude of oscillation of the chemical potential of the left lead.}

\figure{Plot of the normalized amplitude of the right current vs $ \omega $
in the case when the drop  of the potential is inside the 1D wire
(full curve 1) $V_L=V_R=0$, and in the general case for
$V_L/V_R/(EL)=0.15/0.35/0.5$ (dotted line). In the latter case the signal
is approaching its steady behavior (full line 2) with increase of the
frequency.}

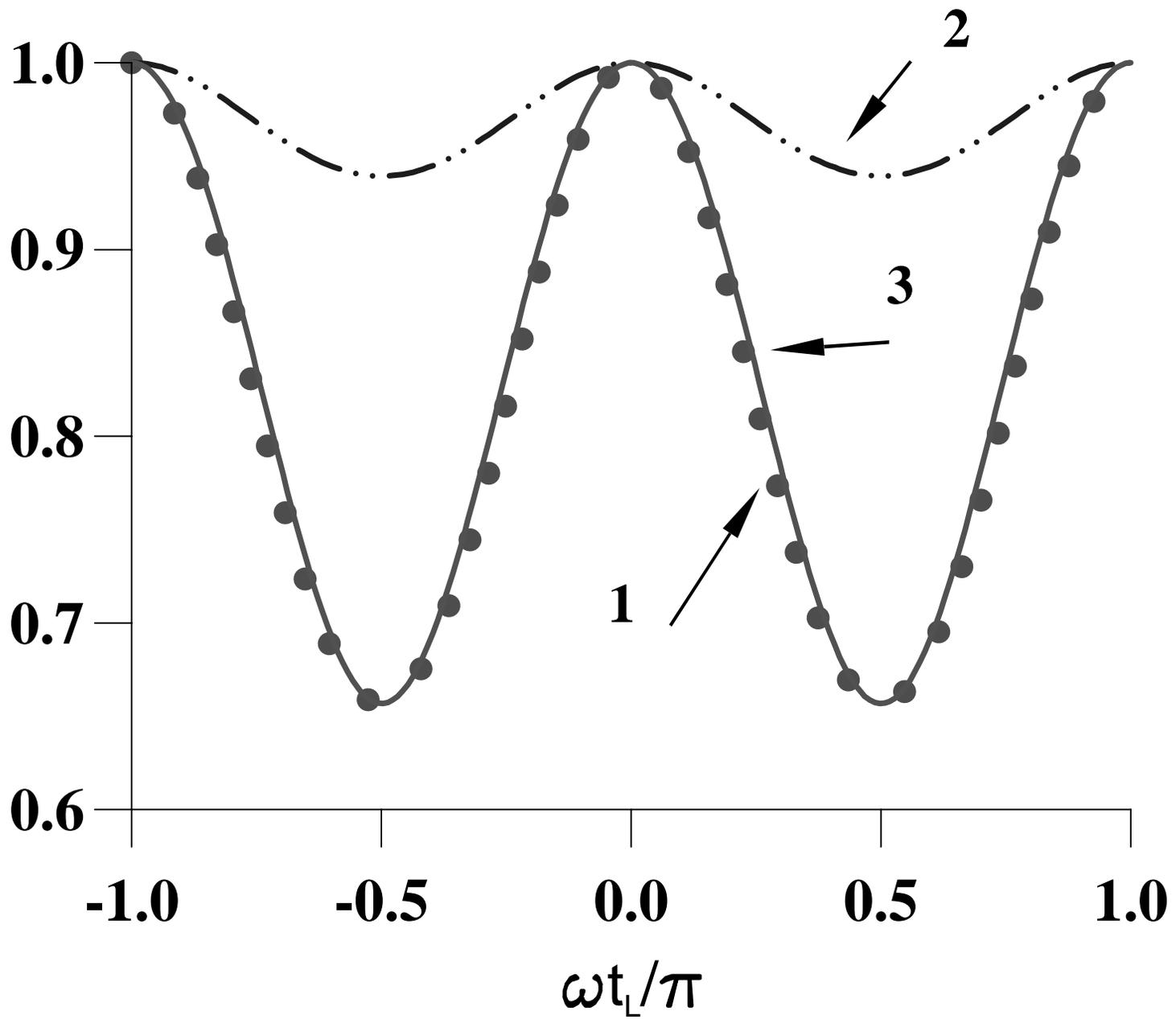
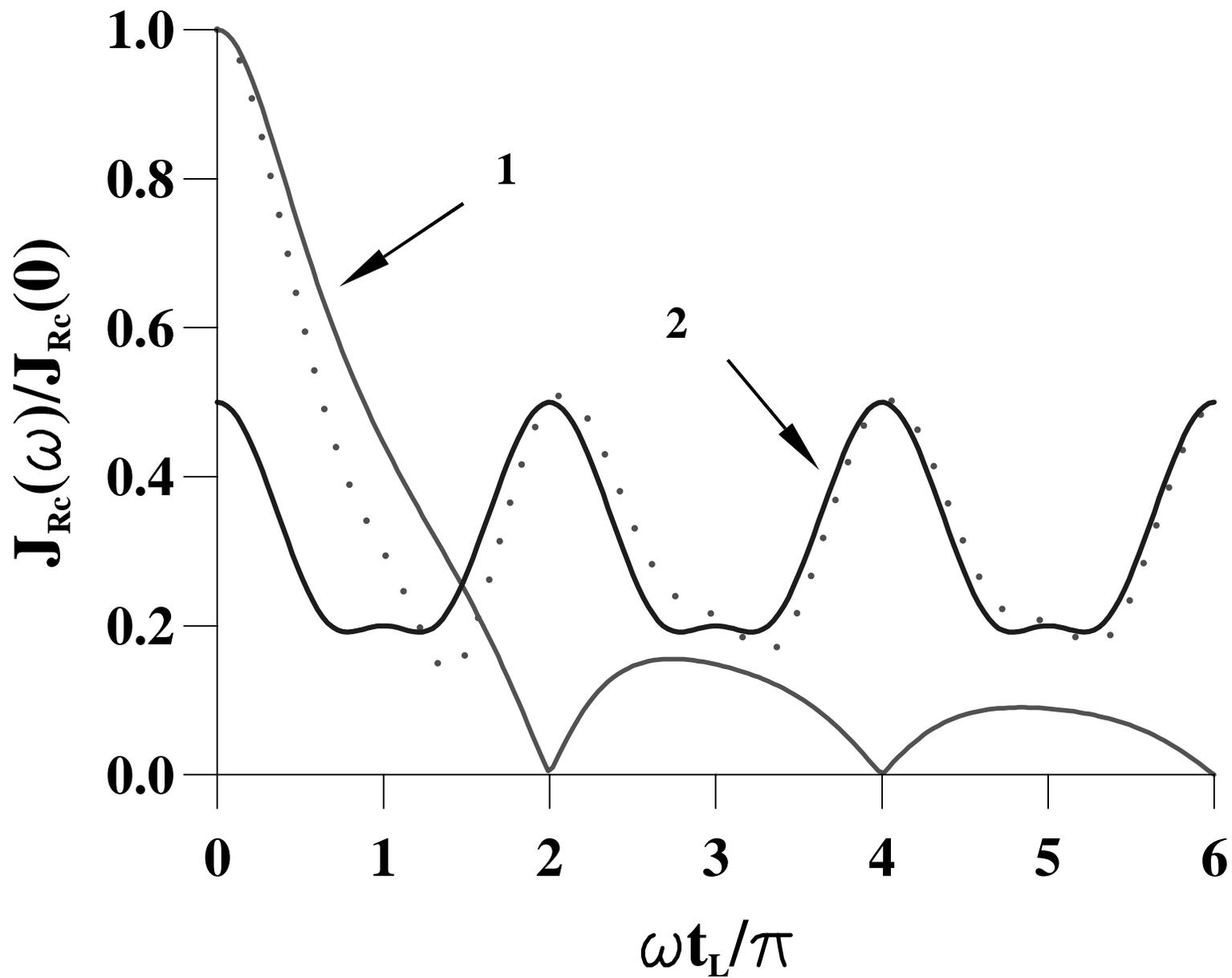
\end{document}